\documentclass{aa}

\usepackage{aabib}
\usepackage{epsfig}
\begin{document}
{

\def\note #1]{{\bf #1]}}
\def\ie{{\rm i.e.}}
\def\muHz{\,\mu{\rm Hz}}
\def\picplace#1{\vspace{#1}}
\def\ddt#1{{\partial #1\over\partial t}}
\def\ddz#1{{\partial #1\over \partial z}}
\def\DDt#1{{{\rm D} #1\over {\rm D} t}}
\def\DD{{\rm D}}
\def\dd{{\rm d}}
\def\vu{\vec{u}}
\def\vxi{\vec{\xi}}
\def\pg{{p_{\rm g}}}
\def\pt{{p_{\rm t}}}
\def\rhobar{{\bar\rho}}
\def\cph{c_{\rm ph}}
\def\ubar{{\bar u}}
\def\uzprime{u_z^\prime}
\def\pbar{{\bar p}}
\def\pgbar{{\bar p}_{\rm g}}
\def\uzbar{{\bar u_z}}
\def\bra{\langle}
\def\ket{\rangle}
\def\Sect#1{Section \ref{#1}}
\def\Eq#1{Eq.\ (\ref{#1})}
\def\Fig#1{Fig.~\ref{#1}}   
\def\apj{{\rm ApJ}}
\def\s{\, {\rm s}}
\def\km{\, {\rm km}}

\newcommand{\FIG}[2]
{
\vspace{0cm}\hspace{0cm}
\epsfxsize=#1 
\epsfbox{#2.eps}
\vspace{0cm}
}

\newcommand{\FIGGIF}[2]
{
\vspace{0cm}\hspace{0cm}
\epsfxsize=#1 
\vspace{0cm}
}

\thesaurus{06(06.09.1;06.15.1)}
\title{Convective contributions to the frequencies of solar oscillations}
\author{Colin S. Rosenthal \inst{1}\thanks{{\it Present Address:\/} Institute Of
Theoretical Astrophysics, P.O. Box 1029, Blindern, Oslo, Norway. colinr@astro.uio.no}
\and
J{\o}rgen Christensen-Dalsgaard \inst{1,2}
\and
\AA ke Nordlund \inst{3,4}
\and
Robert F. Stein \inst{5}
\and
Regner Trampedach\inst{1,2}
}
\institute{Teoretisk Astrofysik Center, Danmarks Grundforskningsfond,
Aarhus Universitet, DK-8000 Aarhus C, Denmark
\and
Institut for Fysik og Astronomi, 
Aarhus Universitet, DK-8000 Aarhus C, Denmark
\and
Niels Bohr Institutet for Astronomi, Fysik og Geofysik,
Astronomisk Observatorium, K\o benhavns Universitet, Denmark
\and
Teoretisk Astrofysik Center, Danmarks Grundforskningsfond, Juliane Maries Vej 30,
DK-2100  K\o benhavn \O, Denmark
\and
Department of Physics and Astronomy, Michigan State University, East Lansing,
MI 48824, USA
}
\offprints{C.S. Rosenthal}
\date{Received $<$date$>$; accepted $<$date$>$}
\maketitle

\begin{abstract}

Differences between observed and theoretical eigenfrequencies of the
Sun have characteristics which identify them as arising predominantly from
properties of the oscillations in the vicinity of the solar surface: in
the super-adiabatic, convective 
boundary layer and above. These frequency differences may therefore provide
useful information about the structure of these regions, precisely where
the theory of solar structure is most uncertain.

In the present work we use numerical simulations of the outer
part of the Sun to quantify the influence of turbulent convection on
solar oscillation frequencies.  Separating the influence into effects
on the mean model and effects on the physics of the modes, we find
that the main model effects are due to the turbulent pressure that 
provides additional support against gravity, and thermal differences 
between average 3-D models and 1-D models.  Surfaces of constant 
pressure in the visible photosphere are elevated by about 150 km, 
relative to a standard envelope model.

  As a result, the turning points of high-frequency modes are
raised, while those of the low-frequency modes remain essentially
unaffected.
The corresponding gradual lowering of the mode frequencies
accounts for most of the frequency difference between observations and
standard solar models.  Additional effects are expected to come primarily
from changes in the physics of the modes, in particular from the modulation
of the turbulent pressure by the oscillations.


\end{abstract}

\section{Introduction}

In standard solar models, the stratification of the convection zone is
determined by mixing-length theory (MLT), thereby reducing the entire
complexity of the turbulent hydrodynamics to a one-parameter family
of models. MLT solar models suffer from several basic inconsistencies.
For example, they predict that convective velocities,
of several km$\,$s$^{-1}$,
disappear abruptly in a few tens of km, immediately below the solar surface.
In contrast,
observations show convective cells with those same characteristic
velocities, and with horizontal sizes of 1000 -- 5000 km, whose velocity
fields obviously cannot vanish so abruptly.
Indirect evidence from spectral
line broadening indeed shows that the photosphere is pervaded by a
velocity field with rms Mach numbers of the order of 0.3, and
yet, in standard solar models these layers are assumed to be entirely
quiescent.  Clearly, such a discrepancy
between the theoretical description and the observations should be
regarded as a warning not to take the quantitative predictions of the
theory too seriously.

Helioseismology provides quantitative diagnostics that pertain precisely
to these critical surface layers, since this is where the upper turning points
of the majority of modes are located.  Thus, analysis of the observed
properties of these modes may help clarify the consequences of the
inconsistencies inherent in MLT and its more recent siblings
(\cite{CM91,CM92,Canuto+Goldman+Mazzitelli96}). Indeed, as we discuss
in more detail below, adiabatic oscillations of MLT models show significant
systematic discrepancies when compared with measured solar frequencies.

It is natural, therefore, to seek to use helioseismology 
applied to these differences to improve the theoretical
description.  However, this procedure is
undermined by our present uncertainty about the physics of the
oscillations near the top of the convection zone where they are likely to
be strongly coupled to both the convective and radiative fields. In the
language of \citetext{Balmforth92b}, the {\emph extrinsic} 
(or {\emph model}\/) error in the mode
frequencies (due to inaccurate modeling of the mean solar structure)
cannot be accurately estimated while the {\emph intrinsic} 
(or {\emph modal}\/) errors
(due to uncertain mode physics) are largely unknown.

One approach to resolving
this problem has been the time-dependent non-local non-adiabatic mixing-length
theory of \citetext{Gough77} and Balmforth (1992abc) in which the
\nocite{Balmforth92a,Balmforth92b,Balmforth92c}
coupling of the oscillations to both the convection and the
radiation is included within the framework of mixing-length theory.
Another approach to the problem has been proposed by
\citetext{Zhugzhda+Stix94} who have developed a model of the modal 
effect on mode
frequencies due to advection of the oscillations by spatially varying
radial flows. This approach has not yet been developed to the stage where
it can be usefully applied to realistic solar models with stratification
and turbulent pressure. Finally, \citetext{Rudiger+97}
have used turbulence
closure assumptions to parameterize the propagation of acoustic
disturbances through a convecting medium.

In the present work, we use an alternative technique for estimating the
model effects, based on the results obtained from
numerical simulations of turbulent convection in a radiating fluid. We show
that p modes can be calculated from a mean model with hydrostatic
and thermodynamic stratification obtained by 
appropriately weighted averages of the simulation results.
We proceed by making simplifying, and certainly very naive, assumptions
about the modal effects, postponing their detailed study to 
subsequent papers.

We begin (Section 2) with a brief discussion of the helioseismic data 
and the discrepancy between measured frequencies and those calculated 
from MLT models.  We then discuss (Section 3) the averaging techniques 
needed to analyze the radial oscillations of a convecting layer, 
describe the model computations (Section 4), and investigate the 
resulting oscillation frequencies (Section 5).  Finally (Section 6),
we discuss the relevance and limitations of the results, and indicate
future plans.

\section{Frequency Residuals for a Standard Solar Model}

\begin{figure}
\FIGGIF{8.8cm}{fig1}
\caption[]{
Measured frequency residuals in the sense (observations) -- (model),
scaled by $Q_{nl}$.
The computed frequencies are for Model~S of \citetext{JCD+96science}.
The symbols indicate
the source of the observed frequencies:
$+$ are LOWL data (\cite{Tomczyk+95}) and $\diamond$ are
data from \citetext{Bachmann+95}
}
\label{data}
\end{figure}

To illustrate the differences between the observed frequencies
and those of current ``standard'' solar models we consider
Model~S of \citetext{JCD+96science}.
This is computed with the OPAL equation of state
(\cite{Rogers+96}),
OPAL opacities (\cite{Iglesias+92}) in
the high-temperature regime and \citetext{Kurucz91} opacities in the
atmosphere.
Diffusion and gravitational settling of helium and heavier elements
are included.
Convection is modeled using \citetext{Boehm-Vitense58} mixing-length
theory. The atmosphere is assumed to be in hydrostatic equilibrium
and is based on a simple analytical $T$--$\tau$ relation,
obtained from a fit to the Harvard-Smithsonian Reference Atmosphere
(\cite{HSRA}).
Evolution starts from
a chemically homogeneous zero-age main-sequence model,
and the model is calibrated to have the current solar
radius and luminosity at an age of $4.6 \, 10^9$years.
The depth of the convection zone in the model of the present
Sun is $0.2885 R_\odot$, close to the helioseismically inferred values
($0.287 R_\odot \pm 0.003 R_\odot$, \cite{JCD+91}; 
 $0.287 R_\odot \pm 0.001 R_\odot$, \cite{Basu+Antia97dCZ}).

In \Fig{data} we show the differences between adiabatic
eigenfrequencies of the standard model and measured solar
p-mode frequencies compiled from the data of
\citetext{Tomczyk+95} and \citetext{Bachmann+95}.
The frequency differences have been scaled with
the quantity $Q_{nl}$ defined as the ratio of the mode mass of
a mode with radial order $n$ and degree $l$ to the mode mass of
a radial mode of the same frequency, the mode mass being defined
by
\begin{equation}
\int_V|\hat{\vxi}|^2\rho_0\dd V= M_{\rm mode}|\hat{\vxi} (R_\odot)|^2 \; ,
\label{modemass}
\end{equation}
where $\hat{\vxi}$ is the displacement eigenfunction of the mode. Here we
have introduced the equilibrium density stratification $\rho_0$ and the
integral is over the volume of the Sun.

Perturbations concentrated in the surface region
of the Sun give rise to frequency residuals which, when scaled with $Q_{nl}$,
are a function of frequency except at high degrees
(e.g. \cite{JCD+Berthomieu91});
it is evident that this is, indeed,
the dominant trend in \Fig{data}. We therefore conclude that the
principal contribution to the solar frequency residuals arises from the
surface layers.  This is consistent with the theoretical arguments that
these layers are particularly badly represented in the models.

Some insight into the effects of this region on the frequencies
might be obtained by linearizing the relation between the structure
and the frequencies in terms of differences between the Sun and a
reference model.
To the extent that the oscillations can be regarded as being adiabatic,
the frequencies are determined by the hydrostatic structure of the
model as well as by the adiabatic exponent
$\Gamma_1 = (\partial \ln p / \partial \ln \rho)_s$
relating pressure $p$ and
density $\rho$, the derivative being at constant specific entropy $s$.
For the purpose of analyzing near-surface effects
a convenient variable is $\upsilon = \Gamma_1/c = 2\omega_{\rm c}/g$,
where $c = (\Gamma_1 p/\rho)^{1/2}$ is the adiabatic sound speed
and $\omega_{\rm c}$ is the isothermal acoustic cutoff frequency.
It was shown by \citetext{JCD+Thompson97}
that if the differences are expressed in terms
of Lagrangian differences ({\ie}, differences at fixed mass)
$\delta_m \upsilon/\upsilon$ and $\delta_m c/c$,
the contribution from $\delta_m c/c$ is small.
Thus the frequency change is approximately linearly related to
$\delta_m \upsilon/\upsilon$,
\begin{equation}
{\delta \omega_{nl} \over \omega_{nl}} \simeq
\int_0^R \tilde K_{\upsilon,c}^{nl}(r) {\delta_m \upsilon \over \upsilon}
{\rm d} r \; ,
\label{freqker}
\end{equation}
where the kernels $\tilde K_{\upsilon,c}^{nl}(r)$ can be calculated
from the structure and eigenfunctions of the reference model.

\begin{figure}
\FIG{8.8cm}{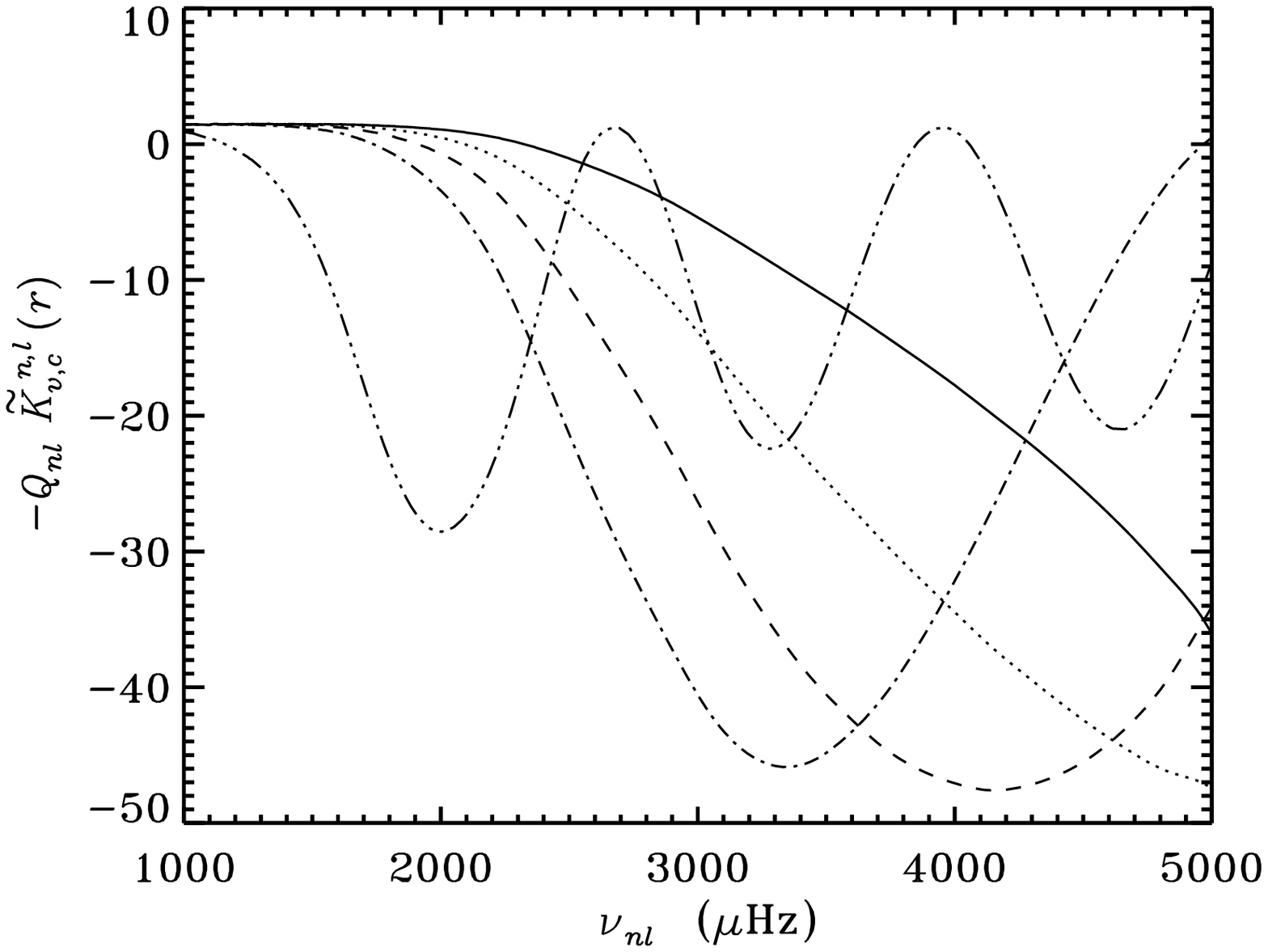}
\caption[]{
Scaled kernels relating the Lagrangian change in $\upsilon \equiv \Gamma_1/c$
to frequency changes (cf. Eq.~\ref{freqker}),
plotted against frequency, for modes of degree $l \le 5$.
Kernels are shown at the photosphere (solid curve)
and at the following depths below the photosphere:
$2\,10^{-4} R$ (dotted curve),
$5\,10^{-4} R$ (dashed curve),
$10^{-3} R$ (dot-dashed curve),
and $5\,10^{-3} R$ (triple-dot-dashed curve)
}
\label{kernels}
\end{figure}

Fig. \ref{kernels} shows $- Q_{nl} \tilde K_{\upsilon,c}^{nl}(r)$
for modes of degree $l \le 5$, plotted against frequency
at several depths;
this corresponds to the effects on the frequencies
of a (negative) delta-function change located at the selected depths.
The scaling with $Q_{nl}$ was included to suppress the effect of
the varying mode mass, whereas the sign is changed to
ease comparison with the differences between observations and
model in \Fig{data}.
Two features of the kernels are immediately evident:
they are very small at low frequency and they become increasingly
oscillatory with increasing depth of the perturbation to the model.
Both features may be understood quite simply in terms of the
properties of the oscillation eigenfunctions.
At low frequencies the upper turning points of the modes are
located at relatively greater depth; hence the eigenfunctions
are evanescent in the region considered and the kernels are therefore small.
At higher frequency the modes are oscillatory quite close to the surface;
as frequency is varied the nodes of the eigenfunction move
relative to the location considered, giving rise to the oscillation,
the shift being more rapid at greater depth.
We note that the observed differences in \Fig{data} are indeed
very small at low frequency and vary slowly with frequency.
This strengthens the conclusion that the major sources of the
differences are located very close to the solar surface.

\section{Averaged Equations for Radial Oscillations}

Our analysis of the equations of momentum and continuity will closely
follow that given by \citetext{Stein+Nordlund91b} in the context of their
analysis of mode excitation. However, the treatment of the energy
equation will require some additional assumptions in order to
obtain a closed form for the resulting equations.

We begin by introducing a horizontal (but not temporal) average denoted
by $ \bra\ldots\ket$ and defining
\begin{equation}
\bar\rho= \bra\rho\ket,\quad \bar p_{\rm g}= \bra p_{\rm g}\ket, \quad {\rm 
and}\quad \bar u_z=
{ \bra \rho u_z \ket\over \bra \rho \ket}
\end{equation}
for the mean density, gas pressure, and vertical velocity. We then define 
fluctuations
around these quantities by
\begin{equation}
\rho'=\rho-\bar\rho\quad{\rm ~etc.}
\end{equation}
The quantities denoted by the overbars thus contain both the mean stratification
and the p-mode oscillations whilst the primed variables are convective
fluctuations. Following  \citetext{Stein+Nordlund91b} it can be
shown that the momentum equation (neglecting dissipative terms) becomes exactly
\begin{equation}
\ddt{\rhobar\ubar}=-{\partial \over \partial z}
\left [ {\rhobar\ubar^2 +  \bra \rho u_z^{\prime 2} \ket } \right ]
-\ddz{\pgbar}+g\rhobar \; .
\label{mom}
\end{equation}
The quantity $\bra \rho u_z^{\prime 2} \ket$ is the turbulent pressure
which we denote by
\def\Pt{\bar p_{\rm t}}
$\Pt$. The total pressure is then defined by
\begin{equation}
{\bar p}={\bar p}_{\rm g}+{\bar p}_{\rm t} \; .
\end{equation}
We now linearize \Eq{mom} by writing
\begin{equation}
\rhobar=\rho_0(z)+\epsilon\rho_1(z,t) \; ,
\end{equation}
with similar expressions for $\bar p$
and $\bar u_z$ with the proviso that $u_{z0}=0$, {\ie}, that there is
no net mass flux through any horizontal surface. Thus we obtain the
linearized momentum equation
\begin{equation}
\rho_0\ddt{u_{z1}}=-\ddz{p_1}+g\rho_1 \; .
\label{linmom}
\end{equation}
Similarly one may show that $\bar\rho$ and $\bar u_z$ satisfy the usual
continuity equation
\begin{equation}
\ddt{\bar\rho}+\ddz{\bar\rho\bar u_z}=0 \; .
\label{contin}
\end{equation}

The energy equation is
\begin{equation}
\DDt{\pg}- {\pg \Gamma_1\over \rho}\DDt{\rho}
=-\left (\Gamma_3-1\right ) \nabla\cdot{\vec{F}}_{\rm rad} \; ,
\end{equation}
where $\Gamma_1$ and $\Gamma_3$ are the usual thermodynamic variables and
${\vec{F}}_{\rm rad}$ is the
radiative heat flux. Averaging as above 
and using equation (9) yields the equation
\begin{eqnarray}
\ddt{\bar \pg}&+&\uzbar\ddz{\bar \pg}+ \bra \pg\Gamma_1 \ket\ddz{\uzbar} = 
\nonumber\\
&&- \bra \pg(\Gamma_1-1){\vec{\nabla}}\cdot{\vec{u}}^\prime \ket
-\ddz{ \bra \pg{\vec{u}}^\prime \ket} \nonumber\\
&&- \bra (\Gamma_3-1){\vec{\nabla}}\cdot{\vec{F}_{\rm rad}} \ket \; .
\label{energy}
\end{eqnarray}

In order to close the system we must make a number of approximations. The
first term on the right-hand side of \Eq{energy} represents the
compressive work and has previously been argued to be negligible
(\cite{Stein+Nordlund91b}). For a perfect gas, dividing $\pg$ by $\Gamma_3 - 1$ 
yields the energy so the second term on the right hand side is the convective flux
divergence multiplied by $(\Gamma_3 - 1)$ while the third term is evidently the 
radiative flux divergence multiplied by the same factor. Their sum is therefore 
proportional to the total flux divergence if departures from a perfect gas law
can be ignored.


In order to proceed with analytical methods
to arrive at an equation for the pulsations, we now
make the simplest possible assumption in \Eq{energy},
namely that the
time-varying part of the sum of the right-hand-side terms may be neglected.
This is obviously just a first, simplified step in a more complete
investigation of the interaction between the convective fluctuations
and the oscillations; even so, it is of interest to investigate
the properties of the oscillations
if the fluctuations of the 3-D averages vanish
and compare the outcome with the observed frequencies.

We note, however, that
the validity of the assumption is clearly suspect since the time scales for
convective motion are comparable with the periods of the oscillations. In the
future we will address the validity of this assumption quantitatively, using
direct measurements of the amplitudes and phases of these terms in
numerical simulations.

In relating the variation in $\pg$ to the variation in total pressure the
two simplest assumptions are
\begin{enumerate}
\item
that the Lagrangian variation in the turbulent
pressure may be neglected, or
\item
that the Lagrangian variation in the turbulent
pressure is directly proportional to the Lagrangian
variation of the gas pressure.
\end{enumerate}
In the first case,
\begin{equation}
\left  \bra \DDt{\pg}\right  \ket=\left  \bra \DDt{p}\right  \ket \; ,
\label{ass}
\end{equation}
leading to the linearized energy equation
\begin{equation}
\ddt{p_1}+u_{z1}\ddz{p_0}
+{\bar \Gamma ^r_1}p_0{\partial u_{z1}\over \partial z}=0 \; ,
\label{linen}
\end{equation}
where we have introduced the so-called ``reduced gamma'' defined by
\begin{equation}
{\bar \Gamma ^r_1}\equiv {  \bra \pg\Gamma_1 \ket_0\over p_0} \; .
\label{rgamma}
\end{equation}
In the second case, if the relative variation is the same in both
components of the pressure, the effective gamma is the
average value $\bra \Gamma_1 \ket_0$ for the gas.

The reduced gamma was originally introduced
by \citetext{Rosenthal+95b}, who argued
that if the Lagrangian perturbation to the turbulent pressure,
$\delta \bar \pt$, were zero then
\begin{equation}
{{\delta \bar p}\over \bar p}={ {\delta\bar \pg} \over \bar p} =
\Gamma_1^r {\delta\bar\rho\over\bar\rho} \; .
\end{equation}
In the present work we investigate the influence of the response of
the turbulent pressure by using two somewhat extreme cases:
the reduced
$\Gamma_1$ (Eq.\ \ref{rgamma}) and average gas $\Gamma_1$. The choice 
of $\Gamma_1$ thus constitutes our
description of the {\emph modal}\/ effects contributing to the measured frequency 
residuals.
The effects of {\emph model}\/ discrepancies are dealt with by using a numerical
simulation of the surface layers to replace the usual MLT prescription, as 
we discuss in detail in the next section.

Equations (\ref{linmom}), (\ref{contin}) and (\ref{linen}), when
supplemented with a prescription for $\Gamma_1$, now constitute a closed
set of equations for the radial oscillations of the convecting layer.
Their form is
identical to the equations for linear, radial oscillations of a
hydrostatic layer
defined by a pressure variation $p_0(z)$ and an adiabatic exponent $\Gamma_1$
given
by either $\bar\Gamma_1^r(z)$ or $ \bra \Gamma_1 \ket_0$.
In order to use results of a simulation to determine the lowest-order
effect on the frequencies of oscillations it is therefore only necessary to
specify these two mean quantities.

We finally note that, even though the derivation presented here
is formally only valid for radial oscillations, we shall
apply the formalism to nonradial oscillations as well;
as in the radial case, this is done by replacing
$\Gamma_1$ in the oscillation equations by
either $\bar\Gamma_1^r(z)$ or $ \bra \Gamma_1 \ket_0$.
For modes of relatively low degree, where the motion is
predominantly vertical in the near-surface layer,
this is likely to be a good approximation,
while it is more questionable for high-degree modes
where the horizontal and vertical components of displacement are comparable.

\section{Model computations}

Our principal goal is to investigate how the treatment of convection
affects the computed structure and oscillation frequencies
in models that cover a large fraction of the Sun.
To achieve this we consider averages of hydrodynamical simulations,
extended continuously with envelope models computed using the
MLT.  We compare these patched models with purely MLT envelopes
using otherwise the same physics,
with the observed frequencies,
as well as with the properties of standard evolutionary solar models
such as Model~S considered in Section~2.

\begin{figure*}
\FIGGIF{17.0cm}{fig3}
\caption[]{
A vertical slice through the simulation showing the temperature
variation. Also shown are the horizontally averaged superadiabatic gradient
and ratio of turbulent to gas pressure
}
\label{simulation}
\end{figure*}
Detailed descriptions of the simulations have been provided
by Stein \& Nordlund (\citeyear{Stein+Nordlund89b,Stein+Nordlund97a})
and \citetext{Nordlund+96bombay}. It is
a 3-dimensional calculation, incorporating LTE radiative transfer and
provisions for using an arbitrary, tabular equation of state.
Most of our results are based on a simulation corresponding to
a rectangular box of horizontal dimensions $6 \times 6$Mm and
a vertical extent of $3.4$ Mm, of which $2.9$ Mm are in the convectively
unstable region.
This was resolved with a mesh with $100\times 100$ horizontal and
82 vertical points, except for a number of tests that were run at
different resolutions (cf.\ Section 5.3 below).

The computation used the MHD equation of state
(\cite{MHD88a,MHD88b,MHD88c,MHD90}),
and an updated version of the \citetext{Kurucz91} opacities
(cf.\ \cite{Trampedach97thesis}).
The chemical composition corresponded to abundances by mass
of hydrogen and heavy elements of $X=0.736945$ and $Z=0.0180550$, respectively;
this is close to the envelope composition of Model~S.
The model assumed a gravitational acceleration of 
$2.740  \, 10^4 \, {\rm cm \, s^{-2}}$ and an average effective
temperature of $5777 \, {\rm K}$.
The averaged models were constructed by averaging first in
the horizontal direction and subsequently carrying out
a Lagrangian time average, {\ie}, a time average on a fixed mass scale,
to filter out the main effect of the p modes
excited in the simulation.
The models were averaged over at least half an hour of solar time.
Figure \ref{simulation} shows a vertical slice through a
single time step of the calculation.
The plot also indicates the averaged superadiabatic temperature
gradient and the ratio of turbulent pressure to gas pressure.
Evidently there is a narrow superadiabatic layer near the top of the box and
the turbulent pressure is significant only near this region.
At greater depths in the box the stratification is nearly adiabatic.

To test the influence of numerical resolution, as well as
of the handling of radiative transfer etc., we have
carried out a number of, generally shorter, simulations with
varying numbers of mesh points (experiments drawn upon in the present work
have resolutions $32^2\times 41$, $50^2\times 82$, $63^2\times 63$,
$100^2\times 82$, $125^2\times 82$, $125^2\times 
163$, and $253^2\times 163$---some of these were calculated with
earlier versions of the equation of state, and have slightly different
chemical compositions).

Since the simulation covers only a narrow region near the top of
the solar convection zone, we require an envelope to which it can be matched.
This was constructed using an MLT envelope code.
It used the same (MHD) equation of state as did the simulation;
also, the
hydrostatic equilibrium in the envelope calculation
included a turbulent pressure of the form
\begin{equation}
\label{pturb}
\pt = \beta \rho v_{\rm con}^2 \; .
\end{equation}
The simulation and envelope models were patched together to produce 
smooth 1-dimensional models. 
This was achieved by adjusting
$\beta$ and the mixing length in the MLT treatment 
such that, at a fixed pressure near the bottom of the simulated
region, turbulent pressure, temperature and density matched continuously.  
The kinetic energy flux was neglected in 
the envelope part of the model; its inclusion would change the 
superadiabaticity of the envelope slightly, but the effect would be 
quite small, since the envelope part of the model is very nearly 
adiabatic.

In addition to being required for the computation of oscillation
frequencies, the extension of the averaged simulations also
provides a measure of the properties of the adiabatic part
of the convection zone corresponding to the simulation.
In particular, it is possible to determine the depth of the
convection zone predicted by the adiabat obtained in the simulation.
Strikingly, the $100^2\times 82$ simulation considered
here leads to a convection-zone depth of $0.286 R_\odot$,
very close to the helioseismically inferred value
($0.287 R_\odot \pm 0.001 R_\odot$, \cite{Basu+Antia97dCZ}).
In view of the fact that no parameters were adjusted to achieve this agreement,
this constitutes additional evidence of the consistency of the model.

It should be noticed that the envelope model does not allow for
convective ``undershoot'' below the convection zone proper.  Thus,
either the undershoot is indeed small, or the agreement of the
convection zone depth is fortuitous.  There is independent evidence,
from simulations of a convection zone with greatly enhanced total
flux (Stein et al. 1997, Dorch 1998) that the convective undershoot is actually small.
\nocite{Dorch98thesis}\nocite{Stein+97insight}

At a total flux $\sim 3 \,10^5$ solar, the convective undershoot
is of the order of 10 -- 20 \% of the solar radius.  The
undershoot may be expected to scale in rough proportion to the
velocity (Hookes law), which again scales roughly as the third root 
of the total flux.  The undershoot at the nominal solar flux is 
therefore expected to be only a fraction of a percent of the solar 
radius.

Errors in the equation of state, or uncertainties in the 
precise abundances of chemical elements could also affect the
apparent match between the predicted and observed depths of the
convection zone.

\subsection{Computation of p-mode frequencies}

Two types of patched models were considered when computing frequencies
of adiabatic oscillations.
In one, in the following referred to as reduced-gamma models (RGM),
the adiabatic exponent between pressure and density was
replaced by the reduced $\Gamma_1$ defined by \Eq{rgamma};
in the second, the gas-gamma models (GGM), the adiabatic
exponent was obtained as the average $\bra \Gamma_1 \ket_0$.
This distinction affects the determination of the adiabatic
sound speed $c$, $\upsilon$ and evidently the oscillation frequencies.
These choices of $\Gamma_1$ are certainly naive, and we do not
expect to obtain a perfect match to the observed frequencies with
any one of them.  However, as stated in the Introduction, in the 
present paper we are mainly concerned with the {\emph model} effects;
i.e., we wish to explore the effects on the oscillation frequencies
that come purely from changes in the mean stratification.  

Pure MLT envelopes were also constructed,
referred to as standard envelope models (SEM).
To isolate the effects of the treatment of convection from other
possible differences in the near-surface region,
they were constructed with the Rosseland opacity 
corresponding to the non-grey opacity used in the simulations.
Furthermore, the atmospheric structure was based on a
$T$--$\tau$ relation obtained as an analytical fit to
the mean $T$--$\tau$ relation for the simulation
(see also \cite{Trampedach+99granada}).
The mixing-length parameter was adjusted
to give exactly the same convection-zone depth as in the
corresponding matched models.

\begin{figure}
\FIG{8.8cm}{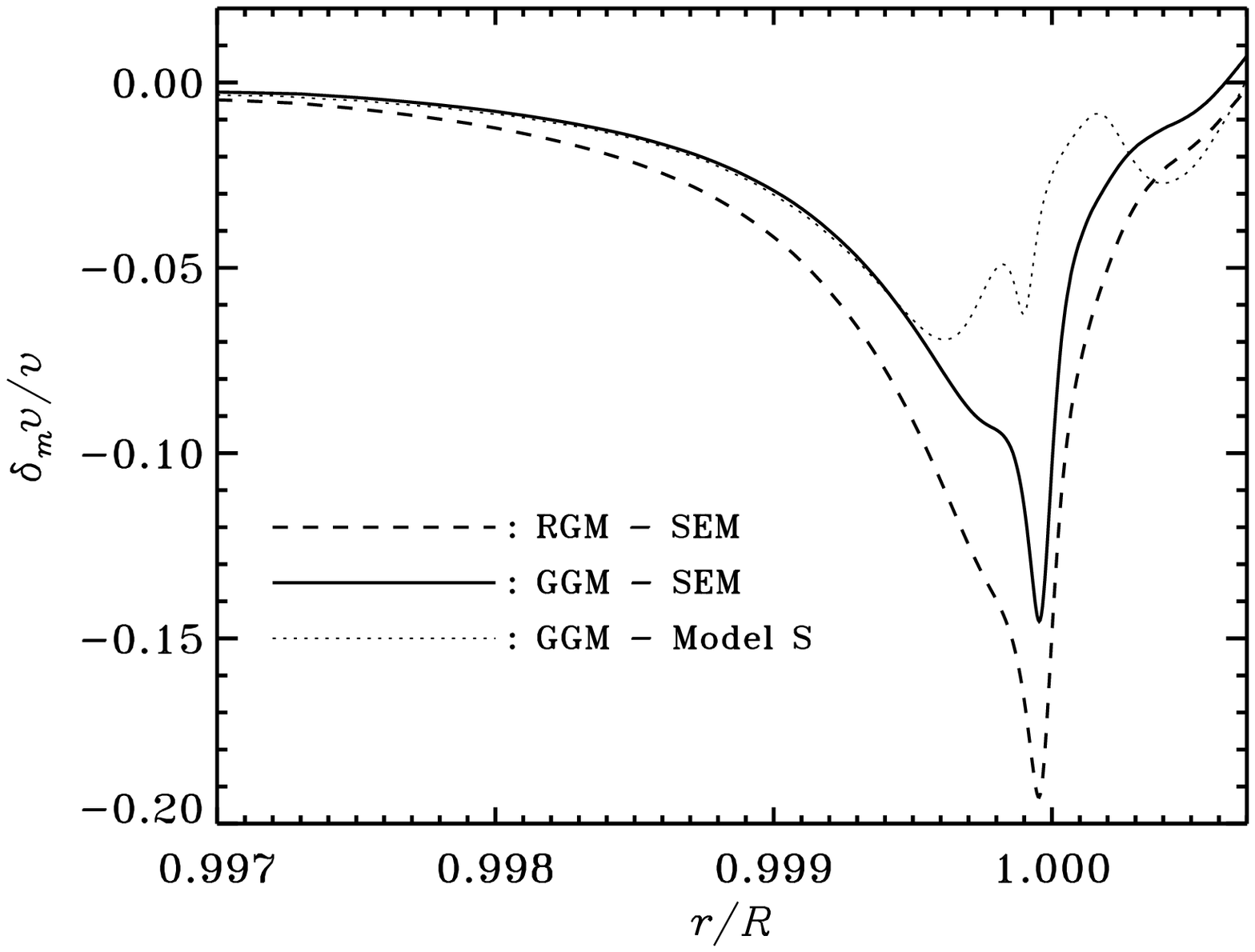}
\caption[]{
Logarithmic Lagrangian (at fixed mass) differences in $\upsilon = \Gamma_1/c$
between patched and comparison envelope models (solid and
dashed lines) and between the GGM envelope and standard solar model
(Model~S)
}
\label{mpdiff}
\end{figure}

Figure \ref{mpdiff} shows the difference in the quantity
$\upsilon=\sqrt{\Gamma_1 \rho_0/p_0}$
between the GGM and SEM and between the RGM and SEM.
The differences were evaluated at fixed mass (or, equivalently, at fixed
hydrostatic pressure).
In the case of the GGM, therefore, the difference largely
reflects the difference in $\rho_0$, caused by the presence of
turbulent pressure and the consequent reduction in the gas pressure;
there is an additional, comparatively small, contribution
arising from the change in $\Gamma_1$ in the region of
steeply increasing hydrogen ionization.
The combined effect is a reduction of $\upsilon$ by up to about 15\%
in a narrow region centered on the superadiabatic layer.
In the RGM, $\upsilon$ is further reduced by the replacement
of $\bra \Gamma_1 \ket_0$ by the reduced $\bar\Gamma_1^r$.
Note that the interior part of the two matched models and the SEM were 
calculated with slightly different mixing-length parameters (as the 
SEM model has no turbulent pressure whereas the matched envelope models
include a turbulent pressure according to \Eq{pturb}), so the models 
are not exactly identical below the matching point.

In addition, the figure shows the difference between the GGM and Model~S.
This difference includes not only effects of turbulent pressure in GGM
but also differences in the $T$--$\tau$ relation and in the atmospheric
opacity.
This is discussed in more detail in section 5.4 below.

\begin{figure}
\FIG{8.8cm}{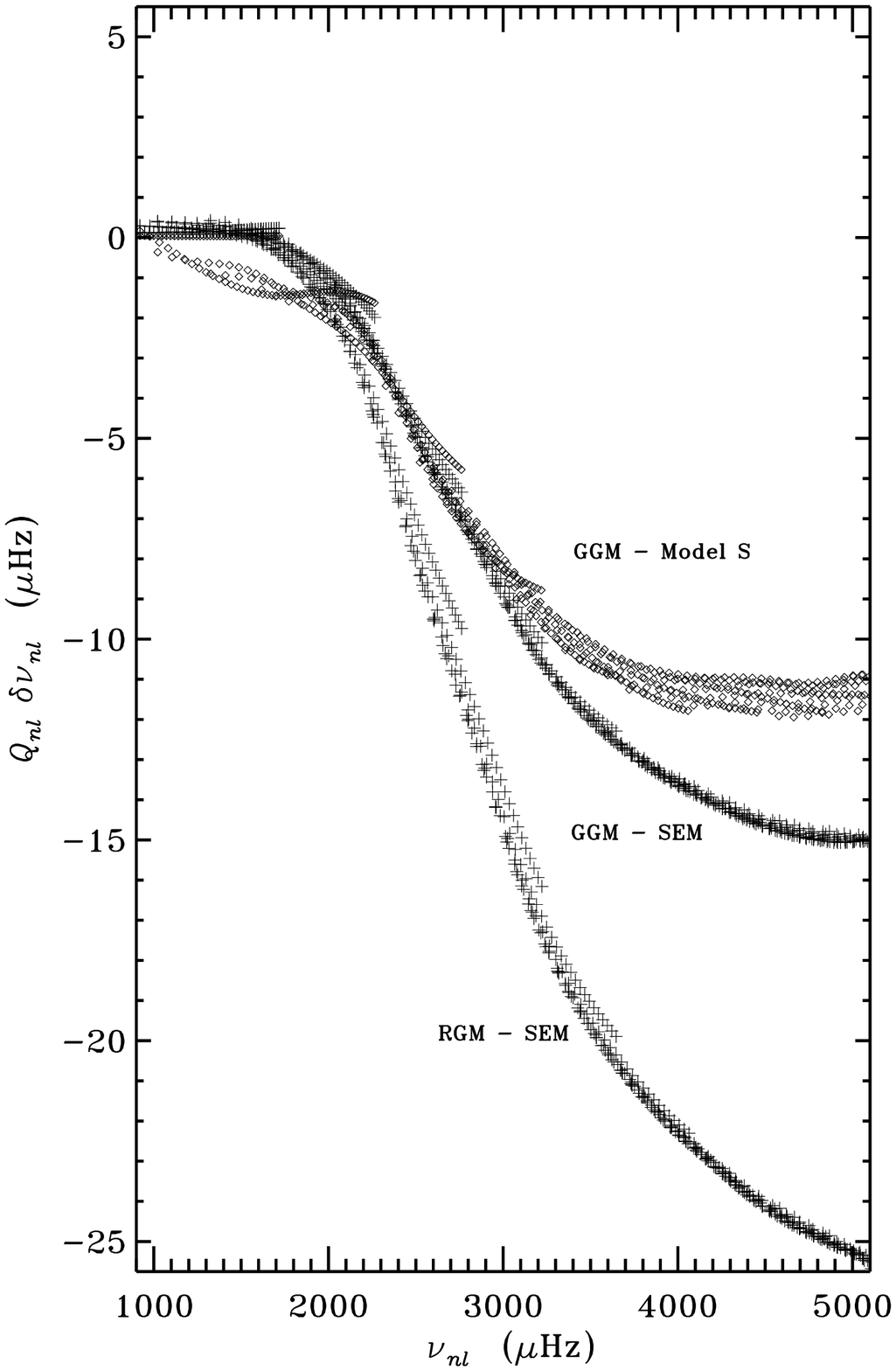}
\caption[]{
Scaled differences
between frequencies of patched and comparison envelope models (crosses)
and between the GGM envelope and standard solar model (Model~S)
(diamonds)
}
\label{mfrdiff}
\end{figure}

\section{Results on oscillation frequencies}

\subsection{Comparison of frequencies}

Adiabatic oscillation frequencies were calculated for the patched
and comparison envelopes, using the procedures described by
\citetext{JCD+Berthomieu91}.
The models are not complete evolved solar
models, unlike  the standard solar model used in
the construction of \Fig{data}. Nevertheless we believe that the
{\emph differences}\/ in mode frequencies between  the models will be
an accurate representation of the actual change in the mode frequencies
of a standard solar model constructed under the same physical assumptions.
In particular, by restricting the comparison to those modes
which are trapped in the convection zone, with $\nu/(l + 1/2) < 50 \muHz$,
the comparison is insensitive to possible differences below the convection zone;
then it is also meaningful to compare the envelope models with
frequencies of Model~S, despite the fact
that due to the inclusion of settling the structure
of the radiative interior of the latter model differs substantially
from the structure of the envelopes.

Scaled frequency differences between the patched and comparison envelopes
are shown in \Fig{mfrdiff}.
To avoid cluttering the diagram only modes with $l \le 300$ were included.
For these, the scaled differences between the envelope frequencies
are indeed predominantly functions of frequency which become
very small at low frequency.
Also, interestingly, the difference between the GGM and SEM frequencies
is of a similar magnitude and shape as the difference between
the observed and Model~S frequencies shown in \Fig{data},
the GGM frequencies being decreased by up to about $15 \muHz$
relative to the SEM frequencies.
On the other hand, in accordance with \Fig{mpdiff}, the
frequency differences for the RGM frequencies are substantially larger.

To interpret \Fig{data}, a more appropriate comparison is
evidently between the frequencies of Model~S and those of the
models including the averaged simulations.
Accordingly, \Fig{mfrdiff} also shows scaled differences
between the frequencies of the GGM and Model~S.
These are slightly smaller than the corresponding differences
between the GGM and SEM; also, they show somewhat larger scatter,
as a result of model differences extending over a larger
part of the convection zone.

\begin{figure}
\FIGGIF{8.8cm}{fig6}
\caption[]{
Measured frequency residuals in the sense (observations) -- (model),
scaled by $Q_{nl}$. The model frequencies are for
the GGM, with the gas $\Gamma_1$.
The symbols indicate the source of the observed frequencies:
$+$ are LOWL data (\cite{Tomczyk+95}) and $\diamond$ are
data from \citetext{Bachmann+95}
}
\label{mobsdiff}
\end{figure}

As argued above, the near agreement  between the depth of the convection
zone in the patched models and in the Sun indicates that the
structure of the deeper parts of the convection zone may be in
similar agreement.
Thus it is of evident interest to compare the computed frequencies
with the observed values.
A comparison of Figs \ref{data} and \ref{mfrdiff}
strongly indicates that the GGM is closer to the solar
data than is the RGM.
Accordingly, in \Fig{mobsdiff} we show scaled differences
between observed frequencies and those of GGM, again restricted
to modes with $\nu/(l + 0.5) < 50 \muHz$ and hence trapped in the
convection zone.
The general magnitude of the p-mode frequency differences
is substantially reduced, and the differences show even less
dependence on degree than in \Fig{data}, indicating that
the origin of the remaining differences is concentrated very close
to the surface.
Note also that differences in \Fig{data}
decrease significantly from zero at rather higher frequency
than do the (GGM) -- (Model~S) differences in \Fig{mfrdiff},
leading to the positive differences around
$3000 \muHz$ in \Fig{mobsdiff}.
As discussed in Section 2 (cf. \Fig{kernels}) this might suggest
that the combined model and modal effects of the differences between the Sun and
Model~S are located somewhat closer to the surface than are
the differences between GGM and Model~S.

The gas $\Gamma_1$ simulation is rather successful in reproducing
the observed mode frequencies.
In contrast, the reduced $\Gamma_1$ gives rise
to a frequency shift which is substantially higher than observed.
Mode physics effects probably account for the remaining discrepancies
in \Fig{mobsdiff}.  These change sign as a function of frequency,
indicating that the effects producing them depend on depth or frequency
or both.  The frequency differences for the f modes, which are barely
affected by changes in the hydrostatic structure of the model or in
$\Gamma_1$, are essentially the same as in \Fig{data} and of similar
magnitude to the other modes now.  Thus, there must be additional mode
physics effects beyond those which can be included in a $\Gamma_1(z,\omega)$.

\subsection{Effects of turbulent pressure}
\label{pturb.sec}

\begin{figure}
\FIG{8.8cm}{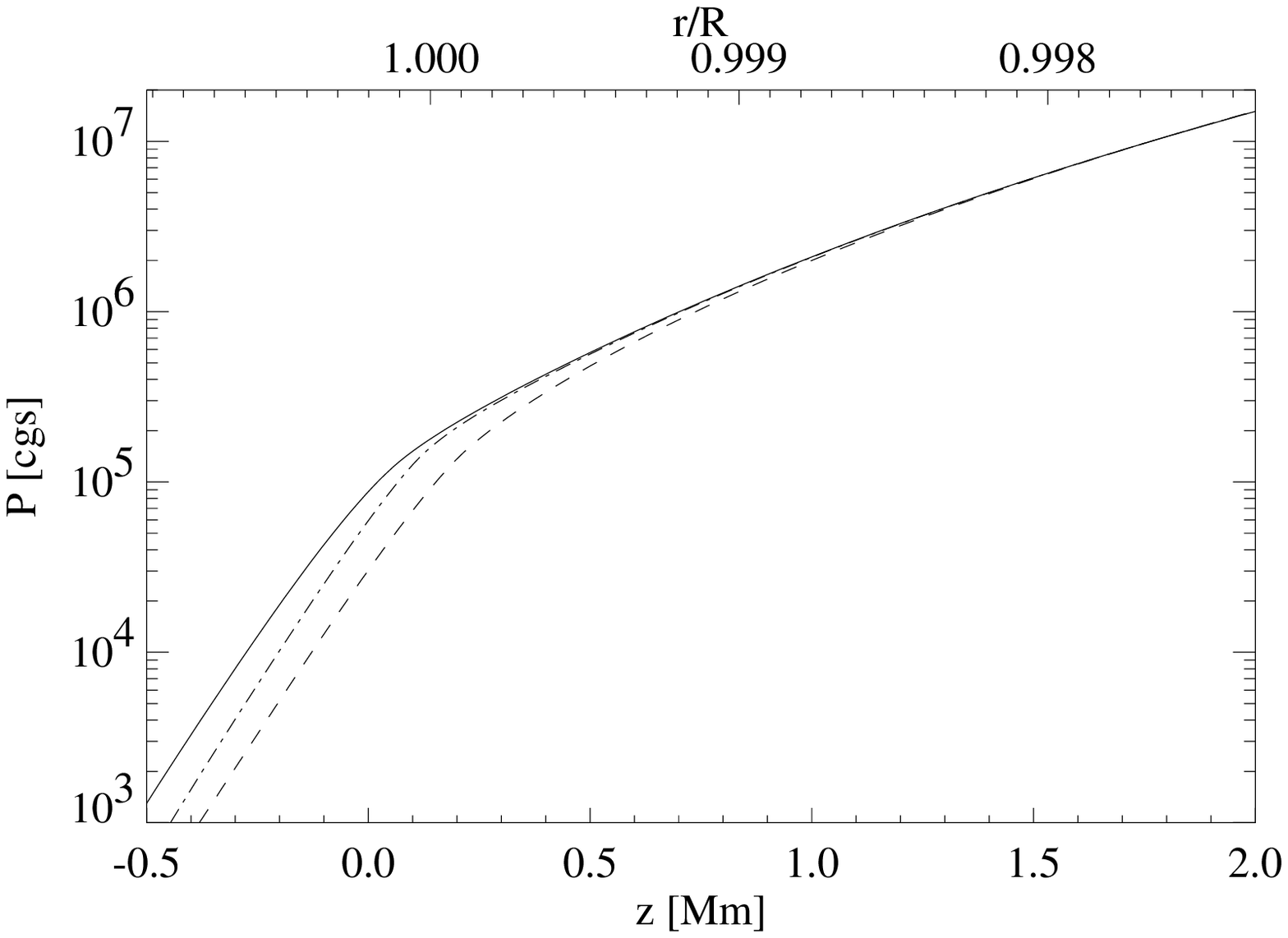}
\caption[]{
Pressure as a function of depth for an averaged 3-D model (full drawn),
and for the comparison standard envelope (SEM) model (dashed).
The dashed-dotted curve shows the pressure stratification of
a 3-D model where the gradient of the turbulent pressure has been 
artificially removed from the vertical pressure balance.
The upper abscissa shows the corresponding position in a complete
model, in terms of the fractional radius $r/R$
}
\label{elev-P.ps}
\end{figure}

The turbulent pressure $\Pt = 
{\bra \rho u_z^{\prime 2} \ket }$ enters the momentum 
equation (\ref{mom}) in an exactly analogous manner to the gas pressure.
It thus adds directly to the gas pressure and provides additional support
against gravity, resulting in an elevation of the solar surface.  The 
extent of the elevation can be determined by comparing the heights of
comparable surfaces of pressure, density or acoustic cutoff frequency in
a normal simulation with one in which the turbulent pressure has been
canceled in the momentum equation.
The result is shown in 
\Fig{elev-P.ps}.  The turbulent elevation of the mid-photosphere
is about 70 km for the $100^2\times82$ model.  Also shown is the
comparison standard envelope (SEM) model.  This model has, by definition,
the same average $T(\tau)$ relation as the 3-D model, and uses the same
opacities, and a conventional mixing length recipe with the mixing
length chosen such as to give the same asymptotic adiabat at depth.

The total elevation of the 3-D model relative to the standard envelope 
model is about 150 km in the mid photosphere, and sets in already 
slightly below the photosphere.  The average temperature in the 3-D 
model is higher than that of the 1-D model in the surface regions, and 
the correspondingly higher pressure scale height causes an elevation 
that adds to the one caused by the turbulent pressure.

The difference in thermal structure reflects fundamental differences 
between 3-D and 1-D models, rather than just differences in the 
efficiency of convective energy transport; a 1-D model with a 
temperature structure consistent with the average pressure 
stratification of the 3-D model would have too high an effective 
temperature, as measured by the emitted surface radiation, while the 
radiation emitted from the 3-D model is consistent with the solar 
effective temperature.  The difference is caused by the temperature 
sensitivity of the opacity; relatively hot regions are more opaque, 
and hence contribute less to the emitted radiation, while relatively 
cold regions are less opaque and hence contribute more.  The higher 
mean temperature of the 3-D model is thus ``hidden'' from view, but is 
reflected in the pressure stratification.

The total photospheric elevation illustrated 
in \Fig{elev-P.ps} provides a simple if crude understanding of
the frequency changes found for the GGM model.
The high-frequency modes have upper turning points in the photosphere;
thus the location of the upper turning point is lifted by the
elevation and hence the acoustic cavity is extended, leading to
a reduction in the frequency.
For radial modes the relative frequency change resulting from this effect
can be estimated as 
\begin{equation}
{\delta \nu \over \nu} \simeq { \Delta r / \cph 
\over \int_0^R \dd r / c} \; ,
\label{bote-dnu}
\end{equation}
where $\Delta r$ is the elevation,
$\cph$ is the photospheric sound speed 
and the denominator is the acoustical radius $\tau_0$ of the star.
Adopting an elevation $\Delta r = 150 \km$,
$\cph = 7 \km \s^{-1}$ and
using $\tau_0 \simeq 3700 \s$, we find 
$\delta \nu / \nu \simeq 6 \, 10^{-3}$,
in general agreement with the frequency changes shown in
\Fig{mfrdiff}.

This argument is roughly consistent with the expression for the
frequency change given in \Eq{freqker}.
Using the asymptotic properties of the eigenfunctions,
neglecting the details of the behaviour near the upper turning point,
\Eq{freqker} may be approximated by
\begin{equation}
\int_0^R {\dd r \over c} {\delta \nu \over \nu} \simeq 
\int_0^R {\delta_m \upsilon \over \upsilon} {\dd r \over c} 
\label{asymp-dnu}
\end{equation}
(\cite{JCD+Thompson97}).
Here, $\upsilon = \Gamma_1 /c = (\rho \Gamma_1 / p)^{1/2}$.
Neglecting the change in $\Gamma_1$,
we find for the right-hand side of \Eq{asymp-dnu}
\begin{eqnarray}
\int_0^R {\delta_m \upsilon \over \upsilon} {\dd r \over c} & \simeq &
{1 \over 2} \int_0^R {\delta_m \rho \over \rho} {\dd r \over c}  \simeq
{1 \over 2} \cph^{-1} \int_0^R {\delta_m \rho \over \rho} \dd r  
\nonumber \\
& \simeq &
{1 \over 2} \cph^{-1} \delta_m r \; ,
\end{eqnarray}
where we took $c \simeq \cph$ to be constant over the region affected by
turbulent pressure and, in the last equality, used
the equation of mass conservation.
Identifying the Lagrangian radius change $\delta_m r$ with $\Delta r$
we recover \Eq{bote-dnu}, apart from the factor $1/2$.

\subsection{Effects of numerical treatment}

\begin{figure}
\FIG{8.8cm}{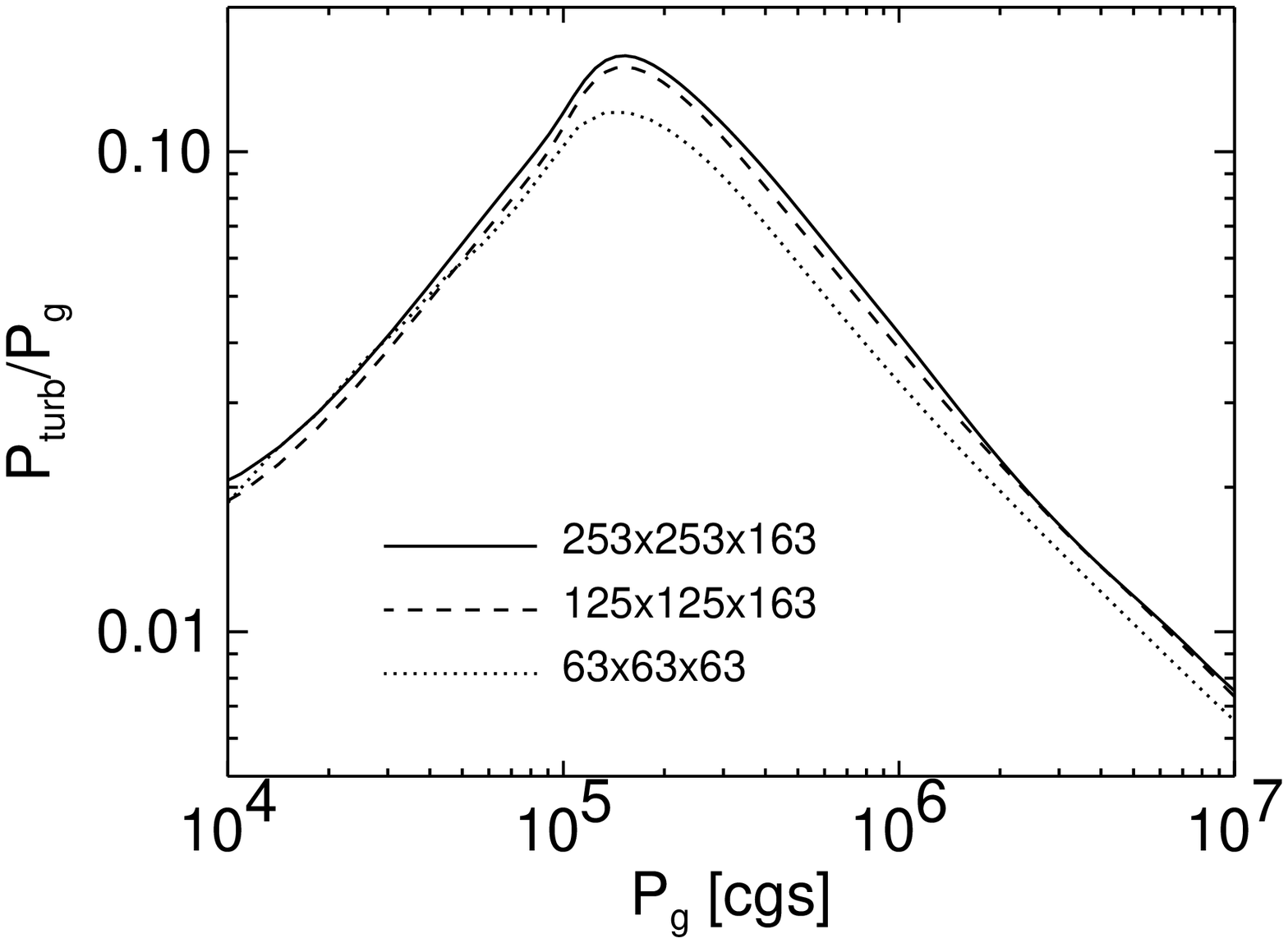}
\caption[]{
Mean turbulent pressure divided by gas pressure as a
function of depth for three different runs, with different numerical
resolutions: $253^2 \times 163$, $125^2 \times 163$, and $63^2 \times 63$. 
}
\label{F8}
\end{figure}

The ratio of turbulent pressure to gas pressure is shown in 
\Fig{F8}.  The figure shows that the shape of the relation is
a very robust property, but that there is an overall scale factor
that depends slightly on the numerical resolution.  The question
thus arises whether the turbulent pressure has converged and whether
it is possible to obtain independent, observational
constraints on its convergence.

The turbulent pressure peaks in a layer that is essentially at the solar 
surface, and the velocity distribution that contributes to the 
turbulent pressure also produces the observed (macro- and micro-) 
turbulent broadening of photospheric spectral lines.  
Figure~\ref{F8} shows that the 
location, shape and width of the maximum in the ratio of turbulent to
total pressure is insensitive to the numerical resolution.  
Horizontal density fluctuations are relatively moderate in the 
visible photosphere, and hence the turbulent pressure $\rho u_{z}^2$
depends mainly on (the square of) the vertical velocity magnitude.

The width of weak iron lines 
observed at solar disc center is a direct measure of the vertical 
velocity amplitudes, and is thus an excellent proxy for the 
turbulent pressure.

Iron lines are ideal diagnostics of photospheric dynamics, not 
only because iron is a sufficiently heavy element for the thermal 
Doppler speed to be smaller than typical photospheric velocities, 
but also because of the large number of iron lines that have 
accurately measured absorption coefficients and wavelengths.  Many 
photospheric lines are blended, but because of the large number of 
iron lines, it is still possible to find dozens of relatively 
clean lines of FeI and FeII in the solar spectrum.

Weak FeII lines have the advantage that they are formed quite close to 
the layer where the ratio of turbulent to gas pressure peaks (cf.\ 
\Fig{F8}), while weak FeI lines are formed about half a pressure 
scale height further up.

\begin{figure}[tbp]
\FIG{8cm}{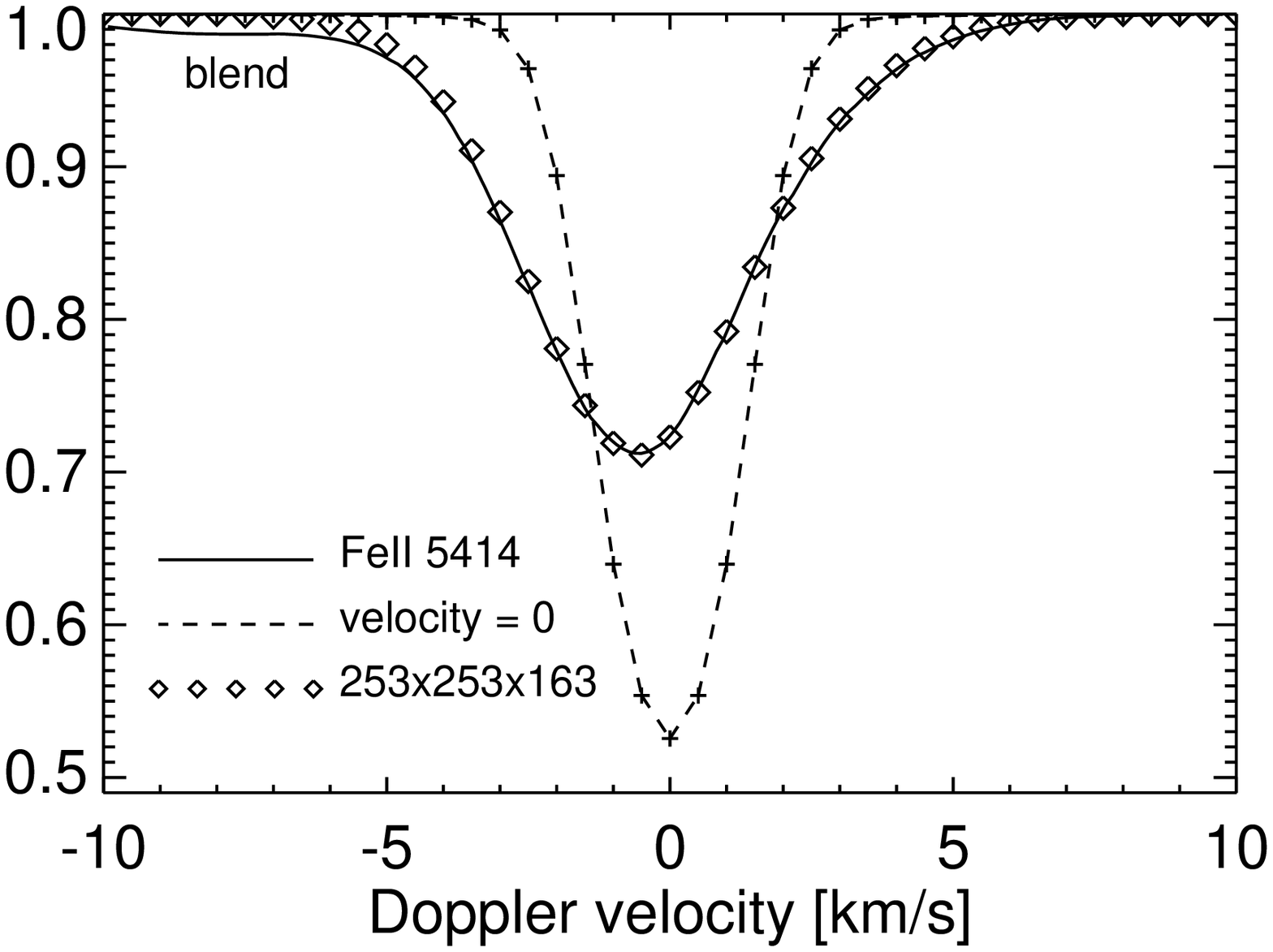}
	\caption[]
{Comparison of observed and synthetic FeII $\lambda5414$ spectral line 
profiles, based on the time average of synthetic spectral line profiles 
from a 253$\times$253$\times$163 simulation.  The observed spectral 
line (from the Li{\'e}ge atlas, \cite{LiegeAtlas73}) is shown full drawn, while the synthetic
spectral line is shown with (diamond) symbols.  A synthetic spectral line
calculated with $u_{z}\equiv 0$ is also shown (dashed line with $+$ symbols).
}
	\label{F9}
\end{figure}
Figure \ref{F9} illustrates the close match that is obtained between 
synthetic and observed weak FeII lines, and also shows what a weak FeII 
line would look like if there were no macroscopic photospheric 
velocities.  The line is much to deep and narrow, and it is perfectly 
symmetric, in contrast to the slightly asymmetric observed spectral 
line.  On the other hand, when the full velocity and temperature 
fluctuations of the 3-D models are included, and a synthetic spectral
line is computed as an average over space and time, the resulting
line matches the width and shape of the observed line closely.  

At a resolution of 253$\times$253$\times$163, the observed line is about 
1.5\% wider than the synthetic spectral line.  Since 
the turbulent pressure scales as the square of the velocity, 
whereas the spectral line width scales linearly with the velocity;
we thus estimate that the turbulent pressure magnitude of the model
is within about 2\% of the correct value.

The similarity of the density stratifications for the various runs 
corroborates this argument.  Even between our lowest and highest 
resolution the change in the elevation would hardly be visible in 
\Fig{elev-P.ps}.  Hence, we expect that the elevation of the 
atmosphere is nearly fully accounted for at our highest resolutions.

In addition to the turbulent pressure stratification, 
the thermal structure is also important in determining the mode frequencies.
The thermal stratification itself is quite insensitive to the numerical 
resolution---there is not much difference even between a model with a 
resolution of only 32$\times$32$\times$41 and one with a resolution of 
253$\times$253$\times$163 (cf.\ \cite{Stein+Nordlund97a}, Fig.\ 26).
The robustness of the average temperature structure is partly due to the 
strong constraints imposed by stratification (Stein \& Nordlund 1989, 
Spruit et al.\ 1990, Spruit 1997, see also Stein \& Nordlund 1998, Fig.\ 
13).

``Spectral line blocking'' plays a critical role in determining the
surface temperature (cf.\ \cite{Mihalas1978}, pp.\ 167-169).  The
blocking of photons by a large number of spectral lines in the solar
spectrum forces an increase of the continuum temperature by about
3\%, relative to a case with no line blocking, in order to maintain 
the solar effective temperature.  The surface temperature in turn,
through the extreme temperature sensitivity of the opacity, 
has a strong influence on the surface pressure.  The temperature
and pressure determine the surface entropy, and are hence of direct
importance for the entropy of the bulk of the convection zone. 

We have made a substantial effort to ensure that the spectral line
blocking is handled as accurately as possible, given the approximations
necessitated by the constraints on computer capacity. During the
simulations, the spectral lines are represented by only four bins
(cf.\ \cite{Nordlund82,Nordlund+Dravins90,Trampedach97thesis}),
but when the lookup tables for the binned opacities and source
functions are computed we ensure, by calibrating against a detailed,
2748 frequency point calculation for a 2-D slice from the model,
that the resulting surface flux is correct, i.e., that the net 
effect of the line blocking is accurately accounted for.  

The accuracy of the spectral line blocking is illustrated by
the close agreement obtained between the predicted depth of the solar 
convection zone and the depth determined from solar observation.
Less complete spectral line data sets (\cite{Bell+76}) that were
used in earlier work resulted in a spectral line blocking that was 
about 40\% smaller.  As a result, the surface temperature was about 
1\% lower, and the solar convection zone was correspondingly about
1.5\% too deep.

The predicted depth of the convection zone depends on series of
compounding factors; the continuum and spectral line opacities
that determine the temperature versus pressure (and hence entropy)
stratification of the surface layers; the convective efficiency of the 
superadiabatic layers immediately below the surface that determines 
the transition to the asymptotic adiabat; and the equation of state
that determines the further run of temperature versus pressure through 
the convection zone.

In this connection it should be noted that the treatment of viscous
and diffusive effects (Stein \& Nordlund 1998, Eqs.\ 4--5) is a matter
of relatively minor importance.  Test runs have shown that reasonable
changes to the numerical coefficients have only minor effects (too small
to be visible in figures such as Fig.\ 26 of Stein \&
Nordlund 1998), and also do not cause visible differences in the
synthetic spectral line profiles (\Fig{F9}).  In effect, such changes
correspond to slight changes in the numerical resolution.  In tests
with driven, isotropic 3-D turbulence, this type of numerical viscosity 
is seen to lead to an inertial range that extends to within about a 
factor of five from the Nyquist wave number, with an average viscous 
dissipation that approaches a limiting value with increasing numerical
resolution.  Since, furthermore, the average viscous dissipation is small 
compared to other terms in the energy equation, the direct influence of 
uncertainties in viscous dissipation is not of great concern.  An 
accurate treatment of the radiative energy transfer, on the other hand, 
is of central importance for the energy balance of the crucial surface 
layers.

Not surprisingly, we conclude that the main factor of importance in  
the numerical treatment is the numerical resolution; the vertical
resolution must be sufficient to resolve the sharp drop in temperature
at the surface of granules, and the horizontal resolution must be
commensurable, to avoid very skew cell aspect ratios.  Ultimately,
whether the numerical resolution is ``sufficient'' or not must be 
judged by empirically studying the convergence of the solution, and
by the consistency of diagnostics such as the synthetic spectral line
profiles shown in \Fig{F9}, the depth of the solar convection zone,
etc..  

Such a procedure can only show consistency, and does not {\em prove} 
that the model is converging towards a correct solution.  
It is 
possible, for example, that the numerical models still contain
inaccuracies that do not significantly 
affect the depth of the convection zone, but which could still have a 
noticeable effect on the surface structure.
However, the more diagnostics that are found to 
be consistent with observations, the smaller is the margin for 
remaining inconsistencies.  The spectral line width data (\Fig{F9}),
for example, does not leave much room for inaccuracies in the 
velocity amplitudes.  Also, since the convective flux depends on 
the product of velocity and temperature fluctuations, errors in
the magnitude of the temperature fluctuations at constant convective
flux would be expected to
lead to errors in spectral line widths.  Spectral line shift and 
bisector agreement (Asplund et al.\ 1999) adds confidence concerning
\nocite{Asplund+99granada} 
the correlation of temperature and velocity fluctuations.

Details could still be improved, though.
A more detailed binning of the opacity could, for example, lead to
a slightly different steepness of the sharp temperature drop at the 
surface, while leaving the overall cooling of gas visiting the surface 
almost unchanged.  It will be necessary carefully to investigate this
and similar issues when studying the more subtle mode physics effects, 
but for the purpose of the present study we are content with the
numerical accuracy -- the overall magnitude of the elevation caused by 
the combination of the turbulent pressure and the thermal differences
is a result that does not depend on such subtleties.

\subsection{Comparison with evolution model}

\label{evol-comp.sec}

We have concentrated on the direct effects of convection
on the model and frequencies, by comparing the averaged
hydrodynamical models (GGM and RGM) with the SEM model
computed with mixing-length theory but otherwise using
as far as possible the same physics as the hydrodynamical models.
However, it is of evident interest to compare also with Model S,
which can be taken as representative of `standard' evolutionary
models of the present Sun.
As seen from \Fig{mpdiff} by comparing the solid and dotted
lines, there are in fact considerable differences between
the SEM and Model S.
We have identified the dominant causes of these differences
as arising from two aspects of the treatment of the atmosphere
of the models: the assumed $T(\tau)$ relation and the
low-temperature opacity.
Both change the temperature structure, as a function of mass,
in the atmosphere and uppermost part of the convection zone;
this in turn affects the density and, in the region of
partial hydrogen ionization, $\Gamma_1$,
leads to the model differences illustrated
(see also Trampedach et al. 1999).
For modes with frequencies below about $3000 \muHz$ 
the upper turning point is essentially below 
the region where these model differences are significant;
hence, in \Fig{mfrdiff}, the frequency differences
for GGM -- Model S and GGM -- SEM are very similar.
At higher frequency, however, the increased model
differences very near the surface, when the SEM model
is used as reference, lead to a significant increase
in the magnitude of the frequency differences.

\section{Discussion}

In the classical theory of solar structure,
a one-parameter family of models (MLT) is
calibrated against the known radius of the Sun.
It is well known that MLT is based on
a number of fundamentally inapplicable and inconsistent assumptions,
and so a large number of alternative models of stellar convection
have been proposed.

Non-local mixing-length theory
(\cite{Gough77};
Balmforth 1992abc\nocite{Balmforth92a,Balmforth92b,Balmforth92c})
attempts to improve on standard MLT
by incorporating into it the effect of the finite size of turbulent eddies.
\citetext{Forestini+91}  have produced an MLT-type model incorporating a
measure of asymmetry between upflows and downflows as found in the
simulations.
The models of \citeauthor{Lydon+92}
(\citeyear{Lydon+93a}, \citeyear{Lydon+93b})
are essentially attempts to parameterize
a wide range of convection simulations using a formalism similar to MLT but
incorporating also the contribution of the kinetic energy flux to the
flux-balance equation. \citetext{CM91} have taken an approach
based on modern theories of turbulence, and have attempted to produce
a parameterized expression based on such theories. Finally,
Canuto (\citeyear{Canuto92,Canuto93})
has produced an ambitious model of convection based on
a Reynolds' stress formalism.

These more elaborate models of convection are potentially very useful,
if it can be shown that they capture essential aspects of the
full 3-D convection in terms of much simpler equations.  In particular,
non-local and time-dependent models of convection such as the ones
by \citetext{Gough77}, Buchler (1993), and Houdek (1997) would be very
\nocite{Houdek97,Buchler93}
useful, for example in the modeling of pulsating stars, since full 3-D 
simulations are too expensive to be used in that context.
The most obvious way to proceed is by use of numerical simulations such 
as the one used here or those of \citetext{Kim+95}, treating the 
simulations as data against which the models are to be tested and
calibrated.

A second approach to testing the simplified models, and one which has
been more widely adopted so far, is the helioseismic approach in which the 
frequencies of
oscillations of a solar model constructed with a given theory are compared with 
the
observed frequencies. Such a procedure, however, must deal with the
difficulty that the complete structure of the surface layers is
certainly underdetermined by the (low- and intermediate-degree) oscillation
frequencies since, as \citetext{JCD+Thompson97} have shown,
the near-surface contribution to the frequencies depends only on $\upsilon$
and not separately on, e.g., $p_0$ and $\Gamma_1$. Thus improved agreement
with measured mode frequencies cannot, by itself, be taken as evidence that a
given model of convection is a better description of reality.

In the present work we have attempted to improve on this approach by
analyzing the problem in such a way as to obtain a physically
justifiable description of the oscillations. Moreover, by using a
numerical simulation of convection we give ourselves no free
parameters with which to calibrate our model. Given this
approach, the fact that we are able to construct a complete solar
envelope model with essentially the correct convection-zone
depth is, in itself, a considerable achievement for the
simulation.

We have here investigated model effects on the mode frequencies,
primarily those due to changes in pressure support of the 
atmosphere and 3-D radiation transfer.  The effects we have found 
are robust; there is no
question that the averaging of 3-D fluctuations results in
differences of this sense and order of magnitude; the turbulent
pressure elevation is constrained by the observed photospheric
velocity field, and the mean thermal difference is an inevitable 
consequence of the temperature dependence of the opacity.

One might attempt to estimate the elevation
effect from turbulent pressure using a local convection model such
as MLT or the model of Canuto \& Mazzitelli (1991, 1992). However,
as discussed by \citetext{Antia+Basu97score}, such models cannot accurately 
account
for the effect of turbulent pressure because they result in an artificially abrupt
upper boundary of the convection zone and therefore a serious overestimate
of the turbulent-pressure gradient there.  In addition, as discussed
in section \ref{pturb.sec}, a 1-D model with the correct pressure stratification
unavoidably has a surface radiation flux that corresponds to an
incorrect effective temperature.  Thus, there are inherent limitations
in simplified 1-D models of convection.

The principal remaining uncertainty in determining the
oscillation frequencies from 3-D models lies in the uncertain mode physics.
In particular, non-adiabatic effects, and the response of the 
turbulent pressure to the compression and rarefaction in the 
oscillations needs to be understood.
We have attempted to quantify this uncertainty by calculating two
models: the RGM in which the turbulent pressure is assumed to be
unaffected by the oscillations and the GGM in which it is assumed to
respond in exactly the same way as does the gas pressure. The frequency
discrepancies for the RGM are almost exactly twice those for the GGM.
The GGM produces frequencies that are closer to those observed,
but this should of course not be taken as evidence that the 
turbulent pressure responds in exactly the same way as the gas
pressure.

The actual depth- and time-dependent mode response of the
turbulent pressure produces, together with the response of the
gas pressure, a complex-valued, and frequency-dependent
$\Gamma_1(z,\omega)$.
At any particular frequency,
the real part of $\Gamma_1$ may be expected to have a different
depth dependence than that assumed in both the RGM and GGM models;
therein lies an essential part of the modal effects,
and thus a potential explanation for part of the remaining differences
between the observed and calculated oscillation frequencies.

Preliminary investigations of the relation between $\delta \ln
\bra \rho \ket$ and $\delta \ln \bra P \ket$ in
numerical 3-D simulations overlaid with initial radial
eigenmodes show that non-adiabatic effects are indeed significant.
The effective gamma appears to be closer to unity than to $5/3$ in
the optically thin parts of the photosphere, while in the very
surface layers the effective $\Gamma_1$ can become quite large
($\sim 8$), because of a localized reduction of $\delta\rho$.
A more quantitative analysis of the non-adiabatic effects will
require much more work, though, and will appear in a subsequent
paper.

Additional differences (in particular the ones reflected in the
discrepancy of the f-mode frequencies) are expected to come from
true 3-D effects; differences between mode behavior in a homogeneous
and inhomogeneous medium.  Again, a quantitative investigation of
such effects requires elaborate, differential comparisons between
non-radial mode behavior in 1-D and 3-D models, and is beyond the
scope of the present paper.

However, a pre-requisite for studying mode physics effects is
a sufficient accuracy of the mean model; only if one includes the 
model effects with sufficient accuracy does it make sense to
use remaining discrepancies to diagnose modal effects.

How accurate, then, is the pressure stratification obtained from the
present numerical simulations?  The tests at various numerical 
resolutions show that the location, shape, and width of the peak of 
the turbulent pressure (relative to the total pressure) are quite
insensitive to the numerical resolution, while the magnitude of
the turbulent pressure increases slightly with increasing numerical
resolution (\Fig{F8}).  The magnitude scaling of the turbulent 
pressure is, however, tightly constrained by spectral line-broadening 
observations (\Fig{F9}).
The existence of turbulent pressure support of about the magnitude 
found here thus cannot be doubted.  Moreover, part of the elevation 
effect is due to the thermal difference between 1-D and average 3-D 
models.  Any calculation of mode frequencies that does not
include a turbulent and thermal pressure elevation of
the upper turning points of the modes is thus neglecting a significant 
effect.  If parameter fitting for such a calculation leads to near 
agreement with the observations it merely illustrates that it is quite 
possible to obtain ``the right result for the wrong reason''.

Finally we must emphasize that while our understanding of convective effects
on radial oscillations may seem rudimentary, it is in fact highly sophisticated
by comparison with our understanding of their influence on nonradial modes.
Indeed if we consider the most nonradial mode of all, the f mode, in
which radial and nonradial motions are of equal magnitude, we note that no
explanation which seeks to replace convection modeling with a hydrostatic
solar model can ever explain the measured frequency residuals because
f-mode frequencies are largely insensitive to hydrostatic structure.
Thus, both the f-mode frequency discrepancies and the remaining
discrepancies in \Fig{mobsdiff} are likely to be caused by modal
effects, rather than by the stratification effects that we have 
uncovered in the present paper.  
Evidently the behaviour of both nonradial and radial 
modes needs more study.  

In order to address the modal effects on nonradial modes 
it will be necessary to
invoke a more elaborate technique than the simple planar averages we have
used, a result already evident from the simplified model calculations
of \citetext{Zhugzhda+Stix94}. On the one hand this emphasizes the gulf
which still exists between theory and measurement in mode physics but,
on a more positive note, it suggests that the remaining frequency 
discrepancies may have great power as diagnostic probes of the 
structure of turbulent convection.

\begin{acknowledgements}
CSR, JC-D, \AA N and RT acknowledge support of Danmarks Grundforskningsfond
through the establishment of the Theoretical Astrophysics Center.
RFS acknowledges support from NASA grant NAG5-4031 and NSF grant 9521785.
\end{acknowledgements}

\bibliographystyle{aabib}
\bibliography{aajour,convection,oscillations,Aake,books,jcd}

}
\end{document}